\documentclass[a4paper]{jpconf}
\usepackage{graphicx}
\usepackage{fancyhdr}
\begin{document}

\def\rar{\rightarrow}
\def\lrar{\leftrightarrow}
\def\mcdot{\!\cdot\!}
\def\erar{&\rightarrow&}
\def\olra{\stackrel{\leftrightarrow}}
\def\ola{\stackrel{\leftarrow}}
\def\ora{\stackrel{\rightarrow}}

\title{Thermal behaviors of light unflavored tensor mesons in the framework of QCD sum rule}

\author{K. Azizi$^{\dag1}$, A. T\"urkan$^{*2}$, H. Sundu$^{*3}$, E. Veli Veliev$^{*4}$, E. Yaz{\i}c{\i}$^{*5}$}

\address{$^{\dag}$Department of Physics, Do\u
gu\c s University, Ac{\i}badem-Kad{\i}k\"oy, 34722 Istanbul, Turkey\\
 $^{*}$Department of Physics, Kocaeli University, 41380 Izmit,
Turkey}

\ead{(Presenter)enis.yazici@kocaeli.edu.tr}

\begin{abstract}
In this study, we investigate the sensitivity of the masses and
decay constants of the light $f_{2}(1270)$ and $a_{2}(1320)$
tensor mesons to the temperature using QCD sum rule approach. In
our calculations, we take into account the additional operators
appearing in operator product expansion at finite temperature. It
is obtained that at deconfinement temperature the decay constants
and masses decrease with amount of $6\%$ and $96\%$ compared to
their vacuum values, respectively. Our results on the masses at
zero temperature are consistent with the vacuum sum rules
predictions as well as the experimental data.
\end{abstract}

\section{Introduction}

In recent years, many researchers are focused on heavy-ion
collision experiments in order to deeply understand the hadronic
dynamics \cite{Apel,Longacre,Doser,Kubota,Ablikim,Aaij}. It is
believed that a transition occurs from hadronic matter to Quark
Gluon Plasma (QGP) phase around critical temperature $T_{c}=175
MeV$. Therefore, thermal QCD calculations on the properties of
hadrons can also be useful in understanding the phase diagram and
other properties of strong interactions. In spite of considerable
theoretical studies on properties of (pseudo)scalar and
(axial)vector mesons at finite temperature
(for instance see \cite{Mallik,Veliev,Veliev3,Loewe,arzu}), there are few theoretical
studies on the thermal properties of the tensor mesons
( as an example see \cite{arzu2}).

In this article, we use the thermal QCD sum rule approach to
investigate the sensitivity of the masses and decay constants of
the light $f_{2}(1270)$ and $a_{2}(1320)$ tensor mesons to
temperature (for the  original work see \cite{arzu1}). In the literature, there are also few studies on the
vacuum properties of tensor mesons using different models
\cite{Aliev,Aliev2,Aliev3,Sundu,Ebert}. The QCD sum rule approach
was firstly proposed by Shifman, Vainshtein and Zakharov
\cite{Shifman} in vacuum as one of the most applicable tools to
hadron physics. This method then was extended to finite
temperature by Bochkarev and Shaposhnikov \cite{Bochkarev}. In
this extension, the Wilson expansion and the quark-hadron duality
approximation are valid, but the quark and gluon condensates are
replaced by their thermal expectation expressions. In thermal QCD
sum rule, due to the breaking of the Lorentz invariance some extra
operators, which are expressed in terms of 4-vector velocity of
the medium and the energy-momentum tensor \cite{Hatsuda,Shuryak}
are appeared in the Wilson expansion. Taking into account these
new operators at finite temperature we obtain the sum rule for the light
tensor mesons under consideration.

\section{Thermal QCD sum rule for light mesons}
In this section our aim is to obtain sum rules for the masses and
the decay constants of the $f_{2}$ and $a_{2}$ tensor mesons. For
this reason, we start with the following thermal correlation
function:

\begin{eqnarray}\label{correl.func.101}
\Pi _{\mu\nu,\alpha\beta}(q,T)=i\int
d^{4}xe^{iq\cdot(x-y)}{\langle} {\cal T}[J _{\mu\nu}(x) \bar
J_{\alpha\beta}(y)]{\rangle}\mid_{y=0},
\end{eqnarray}
where $\cal T$ indicates the time ordering operator and
$J_{\mu\nu}$ is the interpolating current of the tensor mesons.
This current for the mesons under consideration is given by
\begin{eqnarray}\label{tensorcurrent2}
J _{\mu\nu}^{f_2}(x)&=&\frac{i}{2\sqrt{2}}\Big[\bar u(x)
\gamma_{\mu} \olra{\cal D}_{\nu}(x) u(x)+\bar u(x) \gamma_{\nu}
\olra{\cal D}_{\mu}(x) u(x)
\nonumber\\
&+& \bar d(x)\gamma_{\mu} \olra{\cal D}_{\nu}(x) d(x)+ \bar d(x)
\gamma_{\nu} \olra{\cal D}_{\mu}(x) d(x)\Big],
\end{eqnarray}
and
\begin{eqnarray}\label{tensorcurrent2}
J _{\mu\nu}^{a_2}(x)&=&\frac{i}{2\sqrt{2}}\Big[\bar u(x)
\gamma_{\mu} \olra{\cal D}_{\nu}(x) u(x)+\bar u(x) \gamma_{\nu}
\olra{\cal D}_{\mu}(x) u(x)
\nonumber\\
&-& \bar d(x)\gamma_{\mu} \olra{\cal D}_{\nu}(x) d(x)- \bar d(x)
\gamma_{\nu} \olra{\cal D}_{\mu}(x) d(x)\Big],
\end{eqnarray}
where $ \olra{\cal D}_{\mu}(x)$ is the four-derivative with
respect to the space-time acting on the left and right,
simultaneously. We will set $y=0$ after applying derivatives with
respect to $y$.

To obtain the thermal QCD sum rule for the tensor mesons we need to
calculate the correlation function  both in QCD and
hadronic representations. To get the QCD side we use the operator product expansion (OPE) 
to separate the perturbative and non-perturbative contributions. The perturbative part in spectral representation is written as
\begin{eqnarray}\label{QCD Side}
\Pi^{pert}(q,T) =\int ds\frac{\rho(s)}{s-q^2},
\end{eqnarray}
where $\rho(s)$ is the spectral density and it is given by the
imaginary part of the correlation function,
$\rho(s)=\frac{1}{\pi}Im[\Pi^{pert}(s,T)]$. We use the
perturbative parts of the light quark propagator to get the spectral
density (see \cite{arzu1} for more details). It is obtained as
\begin{eqnarray}\label{spectraldenstyf2a2}
\rho_{f_2
(a_2)}(s)=\frac{(m_{u}^{2}+m_{d}^{2})s}{32\pi^2}+\frac{3s^{2}}{160\pi^{2}}.
\end{eqnarray}
 From a similar manner, using the non-perturbative parts of the quark
 propagator we obtain the non-perturbative contributions as:
\begin{eqnarray}\label{nonpertf2a2}
\Pi^{non-pert}_{f_2(a_2)}=\frac{m_d
m_0^{2}}{144q^{2}}\langle\bar{d}d\rangle+\frac{m_u
m_0^{2}}{144q^{2}}\langle\bar{u}u\rangle -\frac{2\langle
u\Theta^{f}u\rangle(q\cdot u)^{2}}{9q^{2}},
\end{eqnarray}
where $\Theta^{f}_{\mu\nu}$ is the fermionic part of the energy
momentum tensor and  $u_{\mu}$ is the four-velocity of the heat
bath. According to the general idea of the QCD sum rules, after
calculation of also the hadronic side of  the correlator we match both
the hadronic and OPE representations of the correlation function. As a result, we obtain the following sum rule:
\begin{eqnarray}\label{rhomatching}
f_{f_2(a_2)}^2(T)m_{f_2(a_2)}^6(T)e^{-m_{f_2(a_2)}(T)/M^{2}}=\int_{(m_u+m_d)^2}^{s_0(T)}
ds \rho_{f_2(a_2)}(s) e^{-s/M^{2}}
+\hat{B}\Pi_{f_2(a_2)}^{non-pert},
\end{eqnarray}
where $\hat  B$ denotes the Borel transformation with respect
to $q^2$, $M^2$ is the Borel mass parameter and $s_0(T)$ is the
temperature-dependent continuum threshold. It is given as
\begin{equation}\label{eqn16}
s_{0}(T)=s_{0}\frac{\langle\bar{q}q\rangle}{\langle0|\bar{q}q|0\rangle}\Big{(}1-\frac{(m_{q}+m_{d})^{2}}{s_{0}}\Big{)}+(m_{q}+m_{d})^{2},  \\
\end{equation}
where $s_{0}$ on the right hand side is the hadronic threshold at
zero temperature.
\section{Numerical Results}
To obtain the values of the masses and decay constants we need to
determine the working regions of two auxiliary parameters: 
the Borel mass parameter and  the hadronic threshold at zero
temperature. We choose the intervals $s_{0}=(2.2-2.5) GeV^2$ and
$s_{0}=(2.4-2.7) GeV^2$ for the continuum thresholds in $f_2$ and
$a_2$ channels, respectively. Also the working region of the Borel mass
is taken as $1.4 GeV^2 \leq M^2\leq 3.0 GeV^2$. In these
intervals, the dependence of the results on the auxiliary
parameters are relatively weak.

Using the working regions of the auxiliary parameters and other input
values, we obtain that the masses and decay constants are well
described by the following fit functions in terms of temperature:
\begin{eqnarray}\label{tetamumu}
m_{f_{2}(a_{2})}(T)==Ae^{\alpha T}+B,
\end{eqnarray}
and
\begin{eqnarray}\label{tetamumu}
f_{f_{2}(a_{2})}(T)=Ce^{\beta T}+D,
\end{eqnarray}
 where the temperature $T$ is in units of GeV. The parameters A,
B, C, D, $\alpha$ and $\beta$ are given in Table 1.
\begin{table}[h]
\renewcommand{\arraystretch}{1.5}
\addtolength{\arraycolsep}{3pt}
$$
\begin{array}{|c|c|c|c|c|c|c|}
\hline \hline  \hline
         &\mbox{A (GeV)} & \mbox{$\alpha (GeV^{-1}$)}& \mbox{B (GeV)} &\mbox{C} &\mbox{$\beta  (GeV^{-1}$)} &\mbox{D}\\
\hline
  \mbox{$f_2(1270)$}        &  -1.055\times10^{-5}   &  72.674  &1.265 &  -4.280\times10^{-6} &  42.808 &  0.043\\
\hline
  \mbox{ $a_{2}(1320)$}        &  -1.255\times10^{-5} & 71.839  & 1.322 &  -4.061\times10^{-6} &  42.662 &  0.042\\

                    \hline \hline  \hline
\end{array}
$$
\caption{Values of parameters A, B, C, D, $\alpha$ and $\beta$ in
fit functions.} \label{tab:lepdecconst}
\renewcommand{\arraystretch}{1}
\addtolength{\arraycolsep}{-1.0pt}
\end{table}

We obtain that the values of the masses and decay constants are
stable until temperature $0.1$  $GeV$,
 but after this point they start to decrease with altering the temperature (see \cite{arzu1} for more details). Also at deconfinement temperature, the decay constants
and masses decrease with amount of $6\%$ and $96\%$ compared to
their vacuum values, respectively.

The obtained results of masses and decay constants at zero
temperature are $m_{f_{2}}=(1.28\pm0.08)~ GeV$,
$f_{f_{2}}=0.041\pm0.002$, $m_{a_{2}}=(1.33\pm0.10)~ GeV$ and
$f_{a_{2}}=0.042\pm0.002$, which are compatible
with the vacuum predictions \cite{Aliev,Aliev2,Ebert} as well as
the existing experimental data \cite{Beringer}. Our predictions on the
temperature behaviors of decay constants and masses can be
verified in the future experiments.

This work has been supported in part by the Scientific and
Technological Research Council of Turkey (TUBITAK) under the
research projects 110T284 and 114F018.

\end{document}